\documentclass[conference]{IEEEtran}
\IEEEoverridecommandlockouts

\usepackage{cite}
\usepackage{amsmath,amssymb,amsfonts}
\usepackage{algorithmic}
\usepackage{graphicx}
\usepackage{textcomp}
\usepackage{xcolor}
\usepackage{booktabs}
\usepackage{multirow}
\usepackage{array}  

\usepackage[top=0.75in, bottom=1.09in, left=0.65in, right=0.65in]{geometry}

\def\BibTeX{{\rm B\kern-.05em{\sc i\kern-.025em b}\kern-.08em
    T\kern-.1667em\lower.7ex\hbox{E}\kern-.125emX}}

\begin{document}

\title{Log2Sig: Frequency-Aware Insider Threat Detection via Multivariate Behavioral Signal Decomposition}

\author{
\IEEEauthorblockN{
Kaichuan Kong\textsuperscript{a}, 
Dongjie Liu\textsuperscript{a,}\IEEEauthorrefmark{1}, 
Xiaobo Jin\textsuperscript{b}, 
Zhiying Li\textsuperscript{a}, 
Guanggang Geng\textsuperscript{a}}
\thanks{\IEEEauthorrefmark{1} Corresponding author. }
\IEEEauthorblockA{
\textsuperscript{a}College of Cyber Security, Jinan University, Guangzhou, China\\
\textsuperscript{b}School of Advanced Technology, Xi’an Jiaotong-Liverpool University, Suzhou, China\\
\ willkkc123@gmail.com, \{djliu, gggeng\}@jnu.edu.cn, tz982354814@163.com, xiaobo.jin@xjtlu.edu.cn \\
}
}

\maketitle

\begin{abstract}

Insider threat detection presents a significant challenge due to the deceptive nature of malicious behaviors, which often resemble legitimate user operations. However, existing approaches typically model system logs as flat event sequences, thereby failing to capture the inherent frequency dynamics and multiscale disturbance patterns embedded in user behavior.
To address these limitations, we propose Log2Sig, a robust anomaly detection framework that transforms user logs into multivariate behavioral frequency signals, introducing a novel representation of user behavior. Log2Sig employs Multivariate Variational Mode Decomposition (MVMD) to extract Intrinsic Mode Functions (IMFs), which reveal behavioral fluctuations across multiple temporal scales.
Based on this, the model further performs joint modeling of behavioral sequences and frequency-decomposed signals: the daily behavior sequences are encoded using a Mamba-based temporal encoder to capture long-term dependencies, while the corresponding frequency components are linearly projected to match the encoder’s output dimension. These dual-view representations are then fused to construct a comprehensive user behavior profile, which is fed into a multilayer perceptron for precise anomaly detection.
Experimental results on the CERT r4.2 and r5.2 datasets demonstrate that Log2Sig significantly outperforms state-of-the-art baselines in both accuracy and F1 score. 

\end{abstract}


\section{Introduction}

Insider threats have emerged as a pressing security issue in enterprise information systems due to their stealthy nature, prolonged attack cycles, and fragmented behavioral patterns. Unlike external attackers, insiders typically possess legitimate credentials and authorized access to internal systems, enabling them to bypass traditional perimeter defenses and camouflage malicious behaviors as routine operations~\cite{alzaabi2024review}. According to the 2025 Ponemon Institute Global Cost of Insider Risk report~\cite{ponemon2025}, organizations encounter an average of 23 insider-related incidents annually, with most attacks taking weeks or even months to detect and contain. These low-frequency, multi-stage, and covert threats present significant challenges to detection mechanisms, particularly in terms of temporal modeling and fine-grained behavioral analysis.

In existing research, insider threat detection is primarily addressed via behavior modeling based on machine learning and deep learning techniques. Traditional machine learning methods extract statistical features, such as login frequency and file access counts, and employ classifiers such as logistic regression (LR), random forest (RF), and XGBoost~\cite{liu2019log2vec, le2020analyzing, bin2022insider} to identify anomalous behaviors. As the sequential and contextual nature of user activities gains increasing attention, deep learning models including LSTM, Transformer, and graph neural networks have been widely adopted for insider threat modeling~\cite{he2021insider, huang2021itdbert, xiao2022robust, roy2024graphch}, thereby improving the ability to capture complex behavior representations and contextual dependencies.

Despite recent advances in modeling accuracy, current approaches still face two key challenges in insider threat detection. \textbf{One challenge lies in modeling behavioral frequency perturbations across activity types}. Insider threats often involve gradual shifts between behavior types, accompanied by evolving frequency patterns. Existing methods based on event counts or discrete sequences struggle to capture such cross-type frequency dynamics, leading to missed threat cues. \textbf{Another challenge is achieving efficient detection over long behavior sequences}. As insider attacks typically span extended time windows, deep models such as Transformers and graph neural networks incur high computational costs when processing long logs, limiting their deployment in latency-sensitive or resource-constrained environments.

To address these challenges, we propose Log2Sig, a novel insider threat detection framework that integrates frequency-aware modeling with efficient sequential representation learning. Log2Sig transforms raw user activity logs into multivariate temporal signals and applies Multivariate Variational Mode Decomposition (MVMD)~\cite{ur2019multivariate} to jointly decompose system-level signals into intrinsic mode functions (IMFs) that reveal behavioral rhythms and multi-scale perturbation patterns. In parallel, we adopt the Mamba~\cite{gu2023mamba} architecture, a structured state space model, as the sequence encoder to capture long-range behavioral dependencies with linear-time complexity, ensuring both strong expressive capacity and deployment efficiency. Finally, we combine the decomposed frequency features with the original event sequences to construct a joint input representation. This enables the model to simultaneously learn temporal dynamics and frequency-domain anomalies, thereby improving detection accuracy and scalability.

Our key contributions are summarized as follows:
\begin{itemize}

    \item \textbf{Log2Sig} is proposed as the first framework to model user activity logs as multivariate frequency signals. By introducing a frequency-aware representation, the framework enables the detection of subtle and multiscale anomalies that are often missed by conventional sequence-based methods.

    \item A frequency decomposition module based on MVMD is developed to extract IMFs across behavioral channels. This approach captures both periodic patterns and non-stationary anomalies, and addresses a core limitation in existing work, which is the inability to model cross-behavioral frequency dynamics.

    \item A dual-view encoding strategy is proposed to model both behavior sequences and frequency-decomposed signals. The former is modeled using a Mamba-based temporal encoder to capture long-range dependencies with linear-time complexity, while the latter is linearly projected to align with the sequence encoding. This design enhances the framework's ability to capture both temporal dynamics and frequency-aware variations over extended time windows.

    \item Extensive experiments on CERT r4.2 and r5.2 datasets demonstrate the superiority of Log2Sig over state-of-the-art baselines. Ablation and robustness studies further confirm the individual contribution of each module and the overall effectiveness of the proposed framework.

\end{itemize}

The remainder of this paper is organized as follows.
Section~\ref{sec:related} reviews related work on insider threat detection and signal-based modeling.
Section~\ref{sec:prelim} introduces the preliminaries of multivariate signal decomposition and the Mamba encoder.
Section~\ref{sec:method} presents the proposed Log2Sig framework, including behavior representation, signal decomposition, encoding, and classification.
Section~\ref{sec:exp} outlines the experimentaldataset,  ssetup, baselines, and evaluation metrics.
Section~\ref{sec:results} reports the empirical results and sensitivity analysis.
Finally, Section~\ref{sec:conclusion} concludes the paper and discusses future directions.

\section{Related Work}
\label{sec:related}
In this section, we review relevant literature in two primary dimensions: (i) insider threat detection based on machine learning and deep learning architectures, and (ii) signal-based methods for behavioral sequence modeling. 

\subsection{Insider Threat Detection}

Insider threat detection has increasingly benefited from machine learning and deep learning approaches, which enable the extraction of temporal, semantic, and structural patterns from user behavior logs. This section presents representative techniques, organized into classical machine learning models and deep learning-based sequential architectures.

\subsubsection{Classical Machine Learning Approaches}

Traditional machine learning methods for insider threat detection primarily rely on discriminative feature extraction from structured audit logs. 
Liu~\emph{et al.}~\cite{liu2019log2vec} proposed Log2vec, a hybrid framework that combines heterogeneous graph embeddings with heuristic rule modeling to capture latent user behavior in enterprise contexts. 
Le~\emph{et al.}~\cite{le2020analyzing} conducted a comparative evaluation of supervised models, including Logistic Regression (LR), Random Forest (RF), and XGBoost, demonstrating that engineered behavioral features can be effectively mapped to risk scores.
Beyond model selection, feature engineering and reduction have proven critical to robustness. 
Bin~\emph{et al.}~\cite{bin2022insider} applied Information Gain (IG) and Correlation-Based Feature Selection (CFS) to eliminate redundancy and improve interpretability. Their findings showed that RF and SVM consistently achieve strong accuracy and generalization across feature subsets.
In unsupervised scenarios where labeled data is limited or unavailable, anomaly detection techniques have gained traction. 
Bartoszewski~\emph{et al.}~\cite{bartoszewski2021anomaly} compared unsupervised models such as Local Outlier Factor (LOF), one-class SVMs (OCSVM), Isolation Forest (IForest), and HMM under both ensemble and single-model settings, emphasizing deployment feasibility using CERT datasets. 
Le~\emph{et al.}~\cite{le2021anomaly} further proposed an autoencoder-based reconstruction method for high-dimensional behavior vectors, while Yousef~\emph{et al.}~\cite{yousef2023machine} employed Isolation Forest to efficiently capture outliers in temporal user logs.

Although these approaches offer strong baselines, their reliance on handcrafted features and limited temporal modeling capacity has motivated the shift toward deep learning.

\subsubsection{Deep Learning-Based Approaches}

Recent advancements in deep learning have enabled more expressive representations of user behavior, improving insider threat detection through modeling of sequential, semantic, and structural dependencies. Early efforts predominantly addressed temporal patterns. 
He \emph{et al.}~\cite{he2021insider} introduced an attention-augmented LSTM framework designed to highlight critical behavioral transitions. 
Building on this, Huang \emph{et al.}~\cite{huang2021itdbert} combined pre-trained BERT embeddings with a bidirectional LSTM to jointly learn contextual semantics and sequential evolution. 
Pal~\cite{pal2023temporal} employed LSTM and GRU networks for temporal representation learning, while Xiao \emph{et al.}~\cite{xiao2024unveiling} integrated CNNs to extract statistical features and Transformers to capture long-range chronological dependencies.
Beyond sequential modeling, representation enhancement techniques have emerged. 
Budžys \emph{et al.}~\cite{budvzys2024deep} proposed GAFMAT, applying Gabor filtering to transform keystroke dynamics into time--frequency representations, thereby improving CNN-based identity modeling. 
Concurrently, Gayathri \emph{et al.}~\cite{gayathri2024spcagan} developed SPCAGAN, a GAN-based framework that generates synthetic insider activity traces via linear manifold learning, mitigating data scarcity in security contexts.
To further capture higher-order relational and structural dependencies, recent works have adopted graph-based paradigms. 
Xiao \emph{et al.}~\cite{xiao2022robust}, Roy \emph{et al.}~\cite{roy2024graphch}, and Cai \emph{et al.}~\cite{cai2024lan} leveraged graph neural networks (GNNs) to jointly model temporal dynamics and inter-user relationships embedded in behavior graphs.

Complementing architectural advances, large language models (LLMs) have recently emerged as a versatile paradigm for log-based anomaly detection. 
LogGPT~\cite{qi2023loggpt} and LogPrompt~\cite{liu2024interpretable} utilize handcrafted prompts for zero-shot or few-shot detection using pre-trained LLMs. 
In contrast, fine-tuning approaches~\cite{song2025confront} adapt LLMs to specific behavioral distributions, enhancing alignment and performance under domain shifts.

Despite the architectural advances and high detection accuracy, many deep learning models suffer from high computational overhead due to complex encoding and training processes. 

\subsection{Signal-Based Modeling for Detection}

Signal decomposition techniques have been applied in various domains, such as wind power forecasting~\cite{yang2024shortterm} and bearing fault diagnosis in mechanical systems~\cite{song2023smart}.However, their use in cybersecurity, particularly for modeling user activity sequences from audit logs, remains underexplored.

Recent efforts have introduced wavelet-based techniques to capture behavioral anomalies. Feng \emph{et al.}~\cite{feng2017wavelet} combined graph-based outlier scoring with Discrete Wavelet Transform (DWT) to detect temporal deviations in user behavior on cloud-sharing platforms. Randive and Ramasundaram~\cite{randive2023mwcapsnet} proposed MWCapsNet, which integrates multi-level 2D wavelet decomposition with capsule networks for image-based behavior modeling, achieving high precision on the CERT dataset. Kim \emph{et al.}~\cite{kim2023anomaly} applied DWT to denoise behavioral features and leveraged fuzzy clustering with OCSVM to reduce false positives. While these approaches benefit from wavelet analysis, they often rely on fixed basis functions and univariate representations.

In contrast, our work employs MVMD~\cite{ur2019multivariate} to decompose behavior frequency signals in a data-adaptive manner, enabling fine-grained and modality-preserving anomaly detection in multichannel activity streams.

\section{Preliminaries}
\label{sec:prelim}
This section briefly reviews two background techniques used in our framework: MVMD for multiscale signal decomposition and Mamba for sequence encoding. We introduce their core principles to support the design of Log2sig.

\begin{figure}[!t]
  \centering

  \begin{minipage}[b]{0.95\linewidth}
    \centering
    \includegraphics[width=\linewidth]{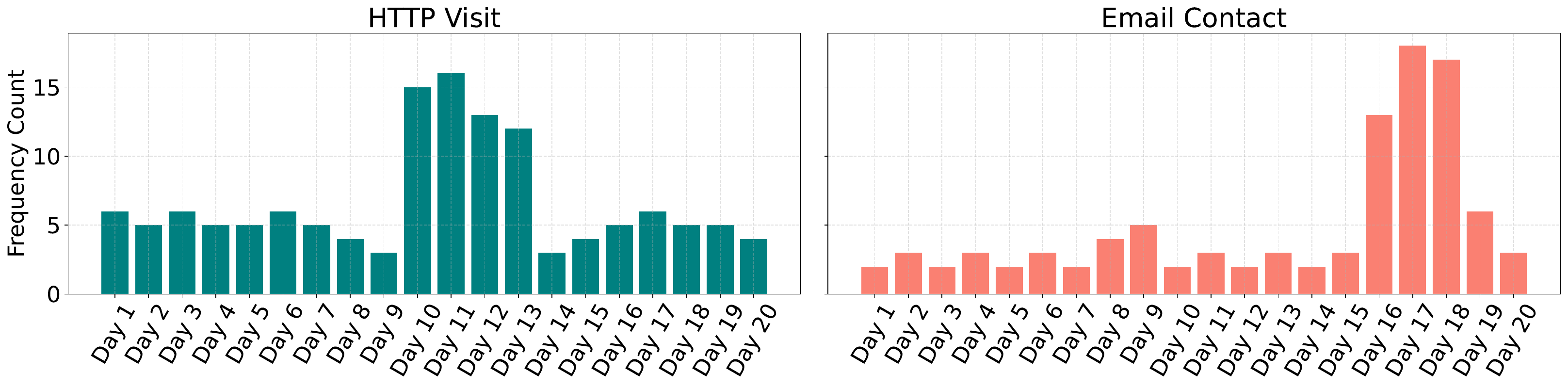}
    \par\vspace{2pt}
    (a) Original Frequency Signal
  \end{minipage}

  \vspace{0.3cm}

  \begin{minipage}[b]{0.95\linewidth}
    \centering
    \includegraphics[width=\linewidth]{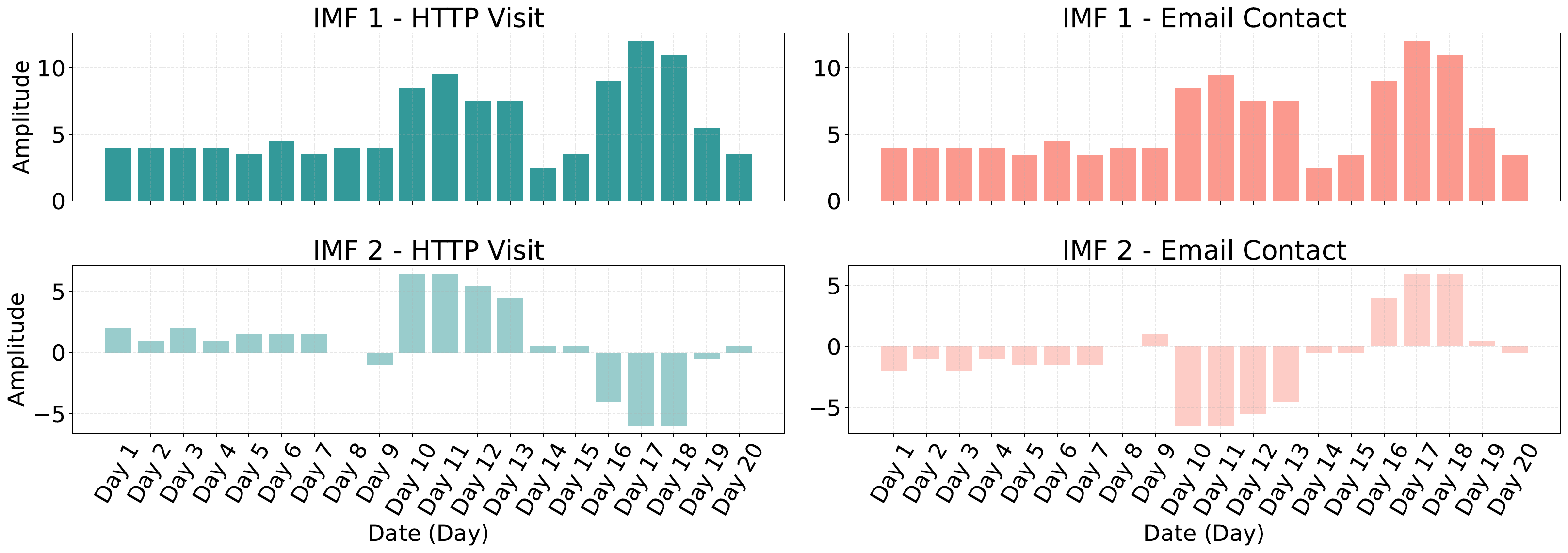}
    \par\vspace{2pt}
    (b) MVMD-Decomposed Frequency Signal
  \end{minipage}

  \caption{Illustrative example of MVMD-based decomposition of user behavior frequency signals.}
  \label{fig:emd_pipeline}
\end{figure}

\begin{figure*}[!ht]
  \centering
  \includegraphics[width=0.9\textwidth]{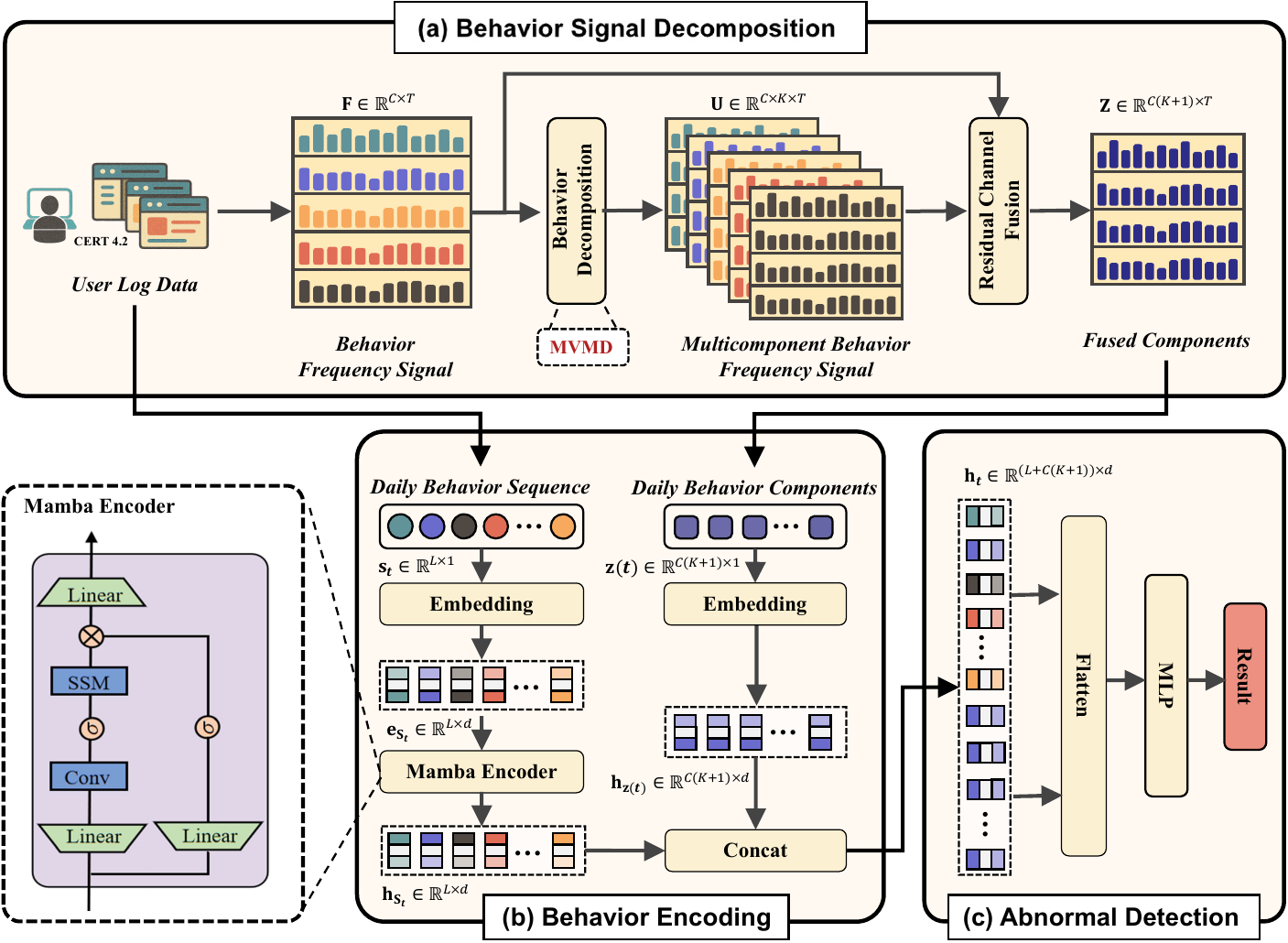}
    \caption{Overview of the Log2Sig framework. (a) User activity logs are transformed into a $C$-channel behavior frequency signal over $T$ days and decomposed by Multivariate Variational Mode Decomposition (MVMD) into multi-scale components. These are fused with the original signal via residual concatenation. (b) In parallel, daily behavior sequences and frequency-based components are embedded and encoded, with temporal patterns captured by a Mamba-based encoder. (c) The combined features are passed through a multi-layer classifier for anomaly detection.}
  \label{fig:Log2sig_framework}
\end{figure*}

\subsection{Multivariate Variational Mode Decomposition}

MVMD~\cite{ur2019multivariate} extends the classical Variational Mode Decomposition (VMD) framework to multichannel settings, enabling joint frequency decomposition across multiple behavioral categories.

Let $\mathbf{y}(t) = [y_1(t), \dots, y_C(t)]^\top \in \mathbb{R}^{C \times 1}$ denote the $C$-channel input signal at time $t$, where each $y_c(t)$ corresponds to the observed activity frequency in the $c$-th behavior category. We adopt a column-vector convention, in which each multivariate observation is stacked channel-wise.

The goal of MVMD is to decompose $\mathbf{y}(t)$ into $K$ multivariate intrinsic mode functions (IMFs) $\{\mathbf{u}_k(t)\}_{k=1}^K$, each capturing a narrowband component with shared spectral structure across all channels. Formally, we have:
\begin{equation}
\mathbf{y}(t) = \sum_{k=1}^{K} \mathbf{u}_k(t), \quad \mathbf{u}_k(t) \in \mathbb{R}^{C \times 1}.
\label{eq:mvmd_sum}
\end{equation}

The decomposition is obtained by solving the following variational optimization problem:
\begin{equation}
\begin{aligned}
\min_{\{u_{k,c}\}, \{\omega_k\}} \quad &
\alpha \sum_{k=1}^{K} \sum_{c=1}^{C} 
\left\| \partial_t \left[ u_{k,c}(t) e^{-j \omega_k t} \right] \right\|_2^2 \\
\text{s.t.} \quad & 
y_c(t) = \sum_{k=1}^{K} u_{k,c}(t),
\end{aligned}
\label{eq:mvmd_loss}
\end{equation}
where $u_{k,c}(t)$ is the $c$-th channel of the $k$-th mode, $\omega_k$ is the center frequency of the $k$-th component, and $\alpha > 0$ is a regularization parameter controlling spectral compactness.

By enforcing spectral alignment across channels, MVMD effectively decomposes multivariate time series into interpretable frequency components. This facilitates robust behavior modeling by capturing both periodic structures and frequency anomalies. The number of modes $K$ and the bandwidth control factor $\alpha$ act as key hyperparameters to adjust decomposition granularity.

To illustrate this decomposition process, Fig.~\ref{fig:emd_pipeline} provides an example on two behavior types: HTTP visits and Email contacts. The extracted IMFs highlight distinct oscillation patterns, separating high-frequency spikes from slower, trend-like variations. This multiscale view facilitates downstream detection of both local anomalies and global drifts.

\subsection{Mamba Encoder}

Mamba~\cite{gu2023mamba} is a structured state space model (SSM) designed for efficient long-range sequence modeling with linear time complexity. Unlike attention-based models, Mamba leverages selective state dynamics and content-aware gating to model temporal dependencies effectively.

Let $\mathbf{X} = [\mathbf{x}_1; \dots; \mathbf{x}_L] \in \mathbb{R}^{L \times d}$ denote the input sequence of $L$ behavior tokens, where each $\mathbf{x}_t \in \mathbb{R}^{1 \times d}$ is a row vector representing the $t$-th token in $d$-dimensional embedding space.

Mamba computes context-aware representations $\mathbf{Z} = [\mathbf{z}_1; \dots; \mathbf{z}_L] \in \mathbb{R}^{L \times d}$ via a selective state-space recurrence defined as:
\begin{equation}
\mathbf{h}_t = \mathbf{h}_{t-1} \mathbf{A} + \hat{\mathbf{x}}_t \mathbf{B}, \quad 
\mathbf{z}_t = \mathbf{h}_t \mathbf{C},
\end{equation}
where $\mathbf{h}_t \in \mathbb{R}^{1 \times d}$ is the hidden state at time step $t$, and $\mathbf{A}, \mathbf{B}, \mathbf{C} \in \mathbb{R}^{d \times d}$ are learnable parameter matrices.

To enable content-dependent modulation, each input token is transformed through a dynamic gating mechanism:
\begin{equation}
\hat{\mathbf{x}}_t = (\mathbf{x}_t \mathbf{W}_u) \odot \sigma(\mathbf{x}_t \mathbf{W}_v),
\end{equation}
where $\mathbf{W}_u, \mathbf{W}_v \in \mathbb{R}^{d \times d}$ are learnable projection matrices, $\sigma(\cdot)$ denotes the sigmoid activation function, and $\odot$ denotes element-wise multiplication. The output $\mathbf{Z} \in \mathbb{R}^{L \times d}$ maintains the original sequence length and embedding dimension.

\section{The Log2sig Framework}
\label{sec:method}

This section introduces the architecture and workflow of the Log2sig framework, which consists of three main components: (1) Behavior Signal Decomposition, (2) Behavior Encoding, and (3) Anomaly Detection. 
Additionally, a preliminary module Behavior Representation Construction is included to standardize raw activity logs into structured behavior inputs. 
An overview of the full framework is depicted in Fig.~\ref{fig:Log2sig_framework}.

\subsection{Behavior Representation Construction}
User activity logs originate from heterogeneous sources such as authentication servers, web proxies, and file access systems, each exhibiting distinct structural formats and semantic conventions. To enable consistent downstream modeling, we employ rule-based mapping strategies~\cite{kong2025dpi} to transform raw logs into a unified schema. Each event is assigned a high-level behavior type (e.g., login, file access), and key attributes—such as timestamps, user identifiers, and action categories—are extracted accordingly.

To preserve both the fine-grained action semantics and the aggregated behavioral statistics, we construct two complementary representations for each user on each day:

\paragraph{Behavior Sequence}  
Let $s_t = [b_1^{(t)}, b_2^{(t)}, \dots, b_L^{(t)}]$ denote the ordered sequence of user $u$’s actions on day $t$, where each $b_i^{(t)} \in \mathcal{B}$ is a token from the behavior vocabulary $\mathcal{B}$. This vocabulary, defined through rule-based aggregation, consolidates heterogeneous events into a compact semantic space. The sequence $s_t$ retains both the temporal and contextual structure of user activities and is subsequently encoded using a Mamba-based sequential model.

\paragraph{Behavior Frequency single}  
In parallel, we compute a daily frequency vector $\mathbf{f}(t) = [f_1(t), f_2(t), \dots, f_C(t)]^\top \in \mathbb{R}^{C \times 1}$, where $f_c(t)$ denotes the count of behavior type $c$ observed on day $t$. Over a span of $T$ days, this forms a multichannel time series:
\begin{equation}
\mathbf{F} = [\mathbf{f}(1), \dots, \mathbf{f}(T)] \in \mathbb{R}^{C \times T},
\end{equation}
where each row traces the temporal evolution of a single behavior category. This structured signal is then passed to the signal decomposition module to extract latent frequency characteristics.

Together, the representation $(s_t, \mathbf{f}(t))$ captures both symbolic action dependencies and quantitative trends, enabling more robust and multiscale behavior modeling.

\subsection{Behavior Signal Decomposition}
While the frequency vector $\mathbf{f}(t)$ provides a compact summary of daily behavior, it lacks the capacity to capture underlying temporal rhythms and frequency-specific patterns. To enrich this representation, we apply MVMD to uncover band-limited components for each behavior type.

\paragraph{Multiscale Behavior Decomposition}
As previously defined, the multichannel behavior frequency signal $\mathbf{F} \in \mathbb{R}^{C \times T}$ consists of daily frequency vectors for $C$ behavior types over $T$ days. We apply MVMD along the time axis to extract $K$ intrinsic mode functions (IMFs) per behavior channel. The result is a three-dimensional tensor:
\begin{equation}
\mathbf{U} = \mathcal{D}_{\text{MVMD}}(\mathbf{F}) \in \mathbb{R}^{C \times K \times T},
\end{equation}
where $\mathbf{U}_{c,k,t}$ denotes the contribution of the $k$-th frequency component for behavior type $c$ at time $t$.

At each time step $t$, the decomposed frequency components are grouped as:
\begin{equation}
\left\{ \mathbf{m}_k(t) = \mathbf{U}_{:,k,t} \in \mathbb{R}^{C \times 1} \right\}_{k=1}^K,
\end{equation}
where each $\mathbf{m}_k(t)$ represents a multichannel behavior vector oscillating at frequency level $k$.

\paragraph{Residual Channel Fusion}
To construct a frequency-enriched representation at each time $t$, we concatenate the original signal $\mathbf{f}(t)$ with all $K$ decomposed components:
\begin{equation}
\mathbf{z}(t) = \text{Concat}[\mathbf{f}(t), \mathbf{m}_1(t), \dots, \mathbf{m}_K(t)] \in \mathbb{R}^{C(K + 1) \times 1}.
\end{equation}

By stacking these residual-enhanced vectors over the entire time window, we obtain:
\begin{equation}
\mathbf{Z} = [\mathbf{z}(1), \dots, \mathbf{z}(T)] \in \mathbb{R}^{C(K + 1) \times T},
\end{equation}
which retains the channel-major layout while embedding rich multiscale frequency information. This representation is then fed into the subsequent encoder for anomaly detection.

\subsection{Behavior Encoding}
To jointly model symbolic behavior sequences and frequency-based statistical patterns, we introduce a dual-view encoding strategy. Each view independently processes one modality, and the resulting embeddings are concatenated to form a comprehensive daily representation. We denote $t$ as the index of the current day in the behavior timeline.

\paragraph{Behavior Sequence Encoding}
The discrete action sequence $s_t = \{b_1^{(t)}, \dots, b_L^{(t)}\}$ is first mapped into a $d$-dimensional embedding space:
\begin{equation}
\mathbf{e}_{s_t} = \text{Embedding}(s_t) \in \mathbb{R}^{L \times d},
\end{equation}
where each token $b_i^{(t)}$ corresponds to a user action occurring on day $t$. The sequence is then passed through a Mamba encoder~\cite{gu2023mamba} to capture fine-grained temporal dynamics:
\begin{equation}
\mathbf{h}_{s_t} = \text{Mamba}(\mathbf{e}_{s_t}) \in \mathbb{R}^{L \times d}.
\end{equation}

Each vector in $\mathbf{h}_{s_t}$ represents a contextualized embedding of the corresponding behavior token.

\paragraph{Behavior Components Encoding}
The frequency-enriched vector $\mathbf{z}(t) \in \mathbb{R}^{C(K+1) \times 1}$, constructed from the original and MVMD-decomposed behavior signals, is treated as a set of pseudo-tokens. Each scalar is projected into the embedding space as:
\begin{equation}
\mathbf{h}_{\mathbf{z}(t)} = \text{Linear}(\mathbf{z}(t)) \in \mathbb{R}^{C(K+1) \times d},
\end{equation}
where each row reflects one behavior-frequency component at day $t$.

\paragraph{Representation Fusion}
To construct the final daily representation, we concatenate the outputs from both encoding branches along the sequence dimension:
\begin{equation}
\mathbf{h}_t = \text{Concat}(\mathbf{h}_{s_t}, \mathbf{h}_{\mathbf{z}(t)}) \in \mathbb{R}^{(L + C(K+1)) \times d}.
\end{equation}
This unified representation integrates symbolic behavior dynamics and frequency-aware statistical patterns, providing a rich embedding for downstream anomaly detection.

\subsection{Anomaly detection}
Based on the fused representation $\mathbf{h}_t$, we employ a lightweight classification module to detect behavioral anomalies on a daily basis.

\paragraph{Representation Flatten}
To enable standard classification, we flatten the representation into a single vector:
\begin{equation}
a_t = \text{Flatten}(\mathbf{h}_t) \in \mathbb{R}^{(L + C(K+1))d \times 1}.
\end{equation}

This transformation preserves both the sequential structure and the multiscale statistics in a high-dimensional feature space. The flattened vector is then passed to a multi-layer perceptron (MLP) classifier:
\begin{equation}
\hat{y}_t = \text{MLP}(a_t), \quad \hat{y}_t \in (0, 1),
\end{equation}
where $\hat{y}_t$ indicates the predicted likelihood of anomalous behavior on day $t$.

\paragraph{Training Objective}
Given a labeled training set $\{(a_t, y_t)\}_{t=1}^T$, where $y_t \in \{0, 1\}$ denotes the ground-truth anomaly label, we optimize the binary cross-entropy loss:
\begin{equation}
\mathcal{L}_{\text{BCE}} = - \frac{1}{T} \sum_{t=1}^T \left[ y_t \log \hat{y}_t + (1 - y_t) \log (1 - \hat{y}_t) \right].
\end{equation}

All parameters in the encoder and classifier are trained end-to-end using the Adam optimizer with appropriate learning rate scheduling.

\section{Experimental Configuration}
\label{sec:exp}
This section describes the datasets, implementation settings, baseline models, and evaluation metrics used in the experiments.

\subsection{Datasets}
We evaluate our method on the publicly available CERT Insider Threat Datasets\cite{lindauer2020insider}, a widely used benchmark for insider threat detection. The r4.2 and r5.2 datasets contain detailed time-stamped logs from 1,000 and 2,000 users respectively, spanning activities such as logon, file access, email, and web usage. Each record is annotated with user IDs and threat labels covering scenarios like data theft and privilege misuse. Dataset statistics are summarized in Table~\ref{table:cert_stats}.

\begin{table}[!t]
    \centering
    \footnotesize
    \caption{Statistics of the CERT Insider Threat Datasets}
    \label{table:cert_stats}
    \begin{tabular}{l|c|c}
        \toprule
        \textbf{Property} & \textbf{CERT r4.2} & \textbf{CERT r5.2} \\
        \midrule
        Time Range & Jan 2010 – May 2011 & Jan 2010 – Jun 2011 \\
        Number of Users & 1,000 & 2,000 \\
        Anomalous Users & 70 & 99 \\
        Total Events & 32,770,222 & 79,856,699 \\
        Anomalous Events & 7,323 & 10,328 \\
        Anomaly Ratio (\%) & 0.022\% & 0.013\% \\
        \bottomrule
    \end{tabular}
\end{table}


\subsection{Implementation Settings}

\begin{table}[!t]
    \centering
    \footnotesize
    \caption{Hyperparameter Configuration of Log2sig}
    \label{table:log2sig_hyperparam}
    \begin{tabular}{p{2.8cm}|p{4.3cm}<{\raggedright\arraybackslash}} 
        \toprule
        \textbf{Component} & \textbf{Configuration} \\
        \midrule
        \multirow{4}{*}{MVMD Decomposition} &  
        Bandwidth $\alpha$: \texttt{500} \\
        &Initialization: \texttt{0} \\
        &Number of Modes $K$: \texttt{3} \\
        &Tolerance: \texttt{1e-3} \\
        \midrule
        \multirow{3}{*}{Mamba Encoder}  &  
        Number of Layers: \texttt{2} \\
        &Embedding Dimension: \texttt{64} \\
        &Normalization: \texttt{RMSNorm} \\
        \midrule
        \multirow{8}{*}{MLP Classifier} &  
        Number of Layers: \texttt{3} \\
        &Hidden Units: \texttt{256-128-32} \\
        &Activation: \texttt{LeakyReLU} \\
        &Dropout: \texttt{0.3} \\
        &Optimizer: \texttt{Adam} \\
        &Learning Rate: \texttt{5e-4} \\
        &Epochs: \texttt{200} \\
        &Batch Size: \texttt{32} \\
        \bottomrule
    \end{tabular}
\end{table}

\begin{table*}[!t]
\centering
\caption{Baseline Performance Comparison}
\label{tab:cert_comparison}
\begin{tabular}{l|cccc|cccc}
\toprule
\multirow{2}{*}{\bfseries Method} & \multicolumn{4}{c}{\bfseries CERT r4.2} & \multicolumn{4}{c}{\bfseries CERT r5.2} \\
\cmidrule(lr){2-5} \cmidrule(lr){6-9}
& Rec  & Prec  & Acc  & F1  & Rec  & Prec  & Acc  & F1  \\
\midrule
IForest         & 0.818 & 0.905 & 0.964 & 0.846 & 0.789 & 0.943 & 0.966 & 0.843 \\
XGBoost         & 0.827 & 0.957 & 0.973 & 0.871 & 0.854 & 0.973 & 0.978 & 0.899 \\
OCSVM           & 0.928 & 0.507 & 0.861 & 0.639 & 0.912 & 0.557 & 0.887 & 0.677    \\
ITDBERT         & 0.884 & 0.912 & 0.960 & 0.898 & 0.889 & 0.914 & 0.961 & 0.901 \\
CATE            & 0.904 & 0.936 & 0.980 & 0.911 & 0.893 & 0.972 & 0.983 & 0.926  \\
LogGPT          & 0.920 & 0.880 & 0.959 & 0.899 & 0.925 & 0.891 & 0.963 & 0.907 \\
ITDLM           & 0.852 & 0.906 & 0.950 & 0.879 & \textbf{0.930} & 0.843 & 0.951 & 0.884  \\
\midrule
\textbf{Log2Sig (Ours)} & \textbf{0.929} & \textbf{0.990} & \textbf{0.990} & \textbf{0.956} & 0.918 & \textbf{0.986} & \textbf{0.988} & \textbf{0.946} \\
\bottomrule
\end{tabular}
\end{table*}

Experiments are conducted on the CERT Insider Threat datasets (r4.2 and r5.2, Scenario 2), focusing on 30 users with verified anomalous behaviors. Daily logs are segmented into behavior sessions comprising behavior sequences and multivariate frequency signals, which are jointly fed into the proposed model. An 80/20 train-test split is employed for evaluation.
To mitigate class imbalance, Synthetic Minority Over-sampling Technique (SMOTE) with a sampling ratio of 0.5 and adaptive neighbor selection is applied, followed by Tomek Links to eliminate borderline instances.

Log2Sig employs a dual-path architecture consisting of a Mamba-based sequence encoder and an MVMD-based frequency decomposition branch. The fused representation is passed through a multi-layer perceptron (MLP) classifier. All modules are trained in an end-to-end manner. Hyperparameter settings are detailed in Table~\ref{table:log2sig_hyperparam}, with tuning procedures discussed in Section~\ref{sec:parameter}.

\subsection{Baselines Methods}
Log2Sig is evaluated against a diverse set of baseline methods, categorized into three major groups:
\textbf{(1) Traditional models}, including IForest~\cite{bartoszewski2021anomaly}, OCSVM~\cite{bartoszewski2021anomaly}, and XGBoost~\cite{le2021anomaly}, which serve as representative unsupervised and supervised learning approaches, respectively;
\textbf{(2) Deep learning models}, such as ITDBERT~\cite{huang2021itdbert} and CATE~\cite{xiao2024unveiling}, which employ Transformer-based or graph-enhanced architectures to capture semantic and structural properties in user logs;
\textbf{(3) Large Language Model (LLM)-based methods}, including LogGPT~\cite{qi2023loggpt} and ITDLM~\cite{song2025confront}, which leverage prompt-driven inference with pretrained LLMs to perform log anomaly detection under zero-shot or few-shot settings.

\subsection{Evaluation Metrics}
Detection performance is assessed using four standard evaluation metrics derived from the confusion matrix: Recall, Precision, Accuracy, and F1-score, defined as follows:
\begin{align}
\text{Recall} &= \frac{TP}{TP + FN} \\
\text{Precision} &= \frac{TP}{TP + FP} \\
\text{Accuracy} &= \frac{TP + TN}{TP + TN + FP + FN} \\
\text{F1-score} &= 2 \times \frac{\text{Precision} \times \text{Recall}}{\text{Precision} + \text{Recall}}
\end{align}

Here, $TP$, $TN$, $FP$, and $FN$ denote the numbers of true positives, true negatives, false positives, and false negatives, respectively. 
\textbf{Recall} quantifies the model’s ability to detect actual threats, while \textbf{Precision} measures the proportion of true threats among all positive predictions. 
\textbf{Accuracy} reflects the overall correctness of predictions. 
\textbf{F1-score} combines Precision and Recall into a single measure, providing a balanced evaluation.

\section{Results and Discussions}
\label{sec:results}
This section reports the experimental results of Log2Sig. We first compare its overall performance against baseline models, then conduct ablation studies to assess key component contributions. We further analyze core hyperparameter sensitivity and evaluate the efficiency of different decomposition and encoding strategies.

\subsection{Baseline Comparison}

\begin{table}[!t]
\footnotesize
\centering
\caption{Impact of Decomposition and Sequence Encoding}
\label{tab:ablation_minimal}
\begin{tabular}{l|cccc}
\toprule
\textbf{Variant} & Rec  & Prec  & Acc  & F1  \\
\midrule
w/o MVMD \& Mamba   & 0.822 & 0.915 & 0.967 & 0.856 \\
w/o MVMD            & 0.833 & 0.979 & 0.977 & 0.890 \\
w/o Mamba           & 0.916 & 0.963 & 0.984 & 0.929 \\
\textbf{Full Model} & \textbf{0.929} & \textbf{0.990} & \textbf{0.990} & \textbf{0.956} \\
\bottomrule
\end{tabular}
\end{table}

As shown in Table~\ref{tab:cert_comparison}, Log2Sig achieves the highest overall performance across both CERT r4.2 and r5.2 datasets. On r4.2, it outperforms all baselines in every metric, achieving an F1-score of 0.956. On r5.2, it maintains strong results with a leading F1-score of 0.946 and slightly lower recall.

Traditional models like IForest and XGBoost perform reasonably but lack adaptability to evolving behavioral dynamics. OCSVM exhibits high recall yet suffers from very low precision, indicating excessive sensitivity to benign anomalies. This stems from its static feature assumptions, which fail to model temporal or structured user behavior.
Deep learning models such as ITDBERT and CATE offer more balanced performance, with CATE benefiting from enhanced structural modeling. However, both rely on static training paradigms that may limit generalization.
LLM-based models (e.g., LogGPT and ITDLM) exhibit improved adaptability to unseen logs. While LogGPT offers balanced precision-recall, its effectiveness diminishes under dynamic conditions.

In contrast, Log2Sig integrates multiscale frequency decomposition with sequence modeling, enabling robust, high-precision detection of subtle threats. These results demonstrate its superiority in both static and dynamic insider threat scenarios.

\begin{figure*}[!t]
  \centering

  \begin{minipage}[b]{0.43\linewidth}
    \centering
    \includegraphics[width=\linewidth]{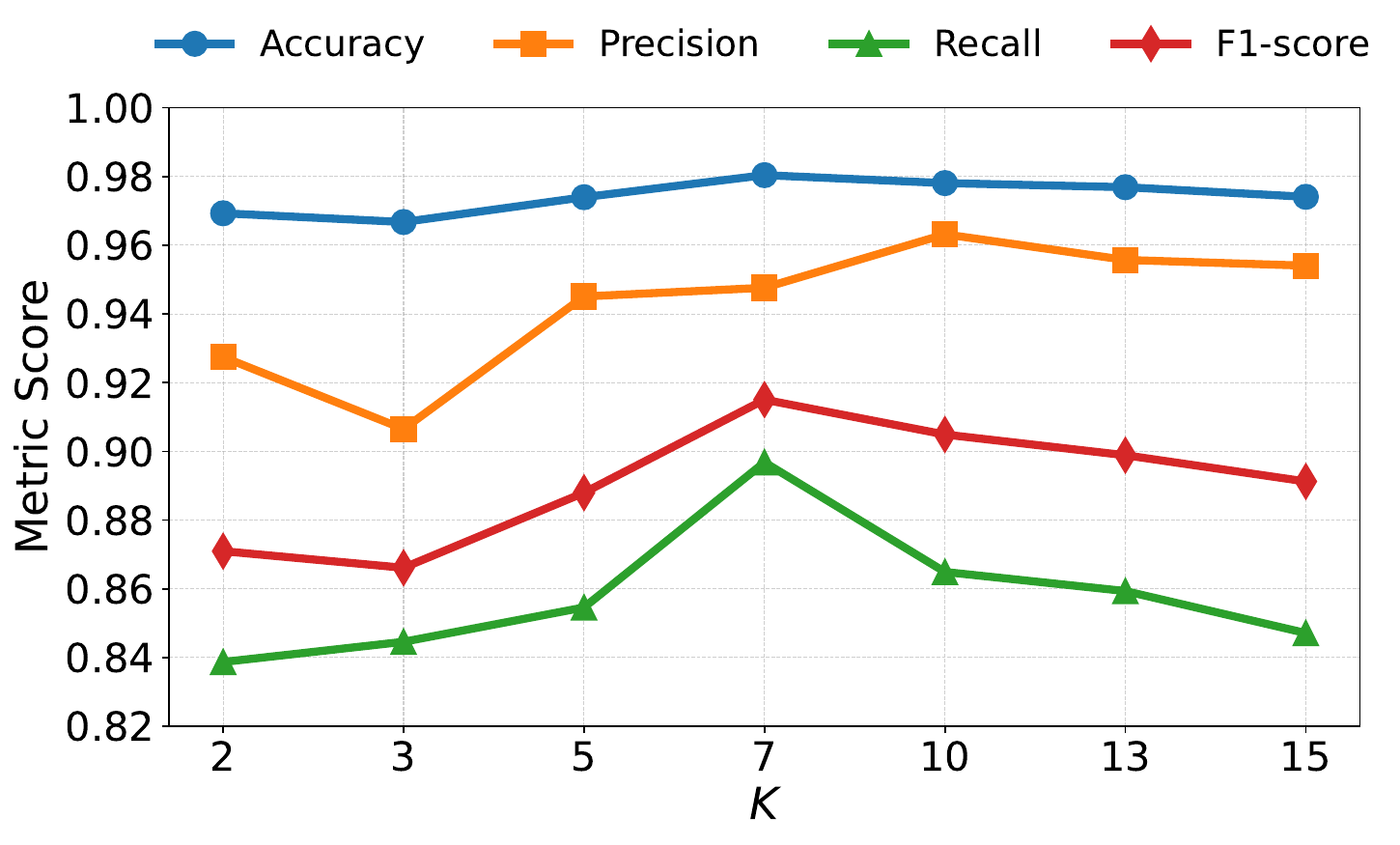}
    \par\vspace{2pt}
    (a) Effect of MVMD Mode Number $K$
  \end{minipage}
  \hspace{0.05\linewidth}
  \begin{minipage}[b]{0.43\linewidth}
    \centering
    \includegraphics[width=\linewidth]{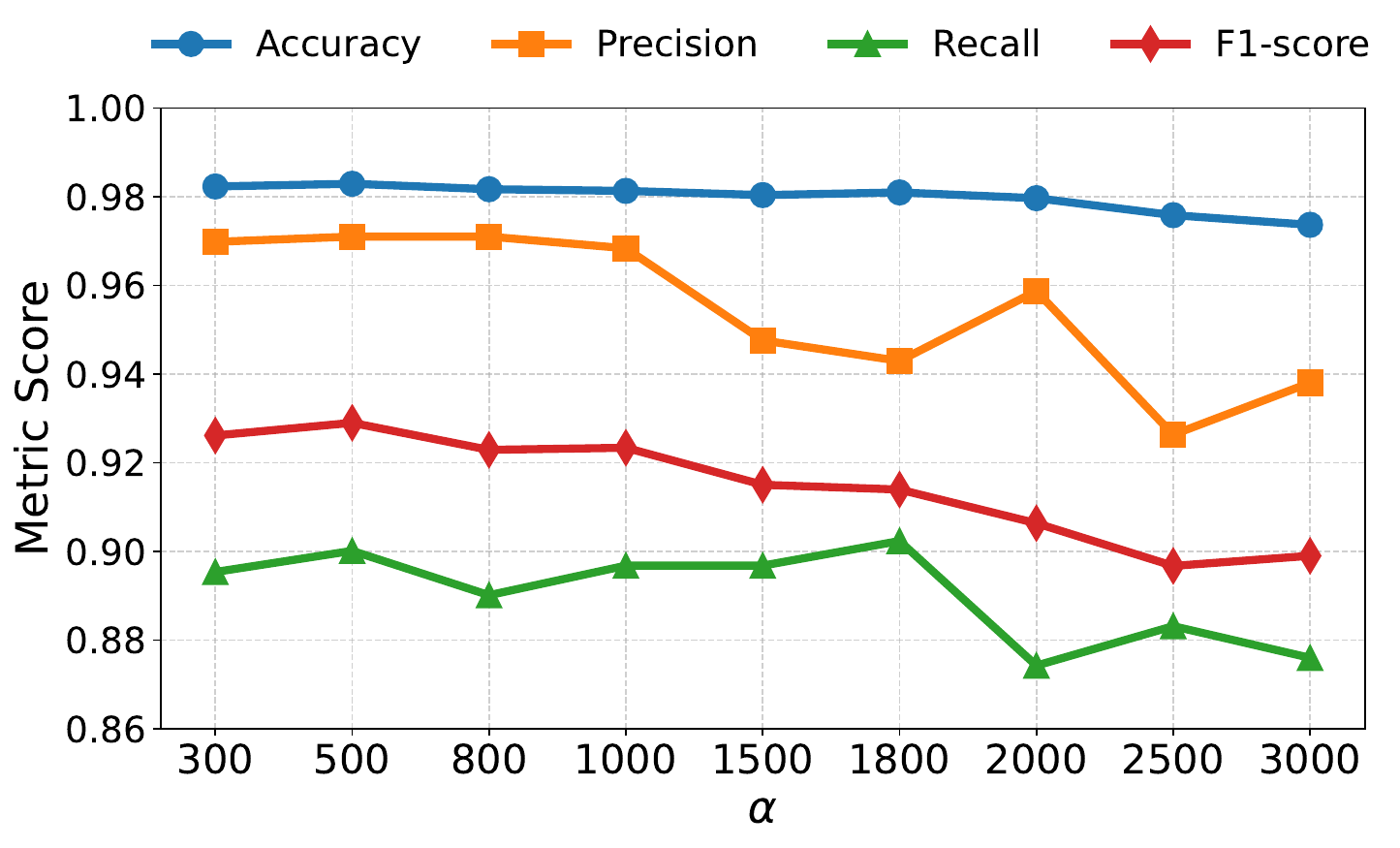}
    \par\vspace{2pt}
    (b) Effect of MVMD Bandwidth Parameter $\alpha$
  \end{minipage}

  \vspace{0.2cm}

  \begin{minipage}[b]{0.43\linewidth}
    \centering
    \includegraphics[width=\linewidth]{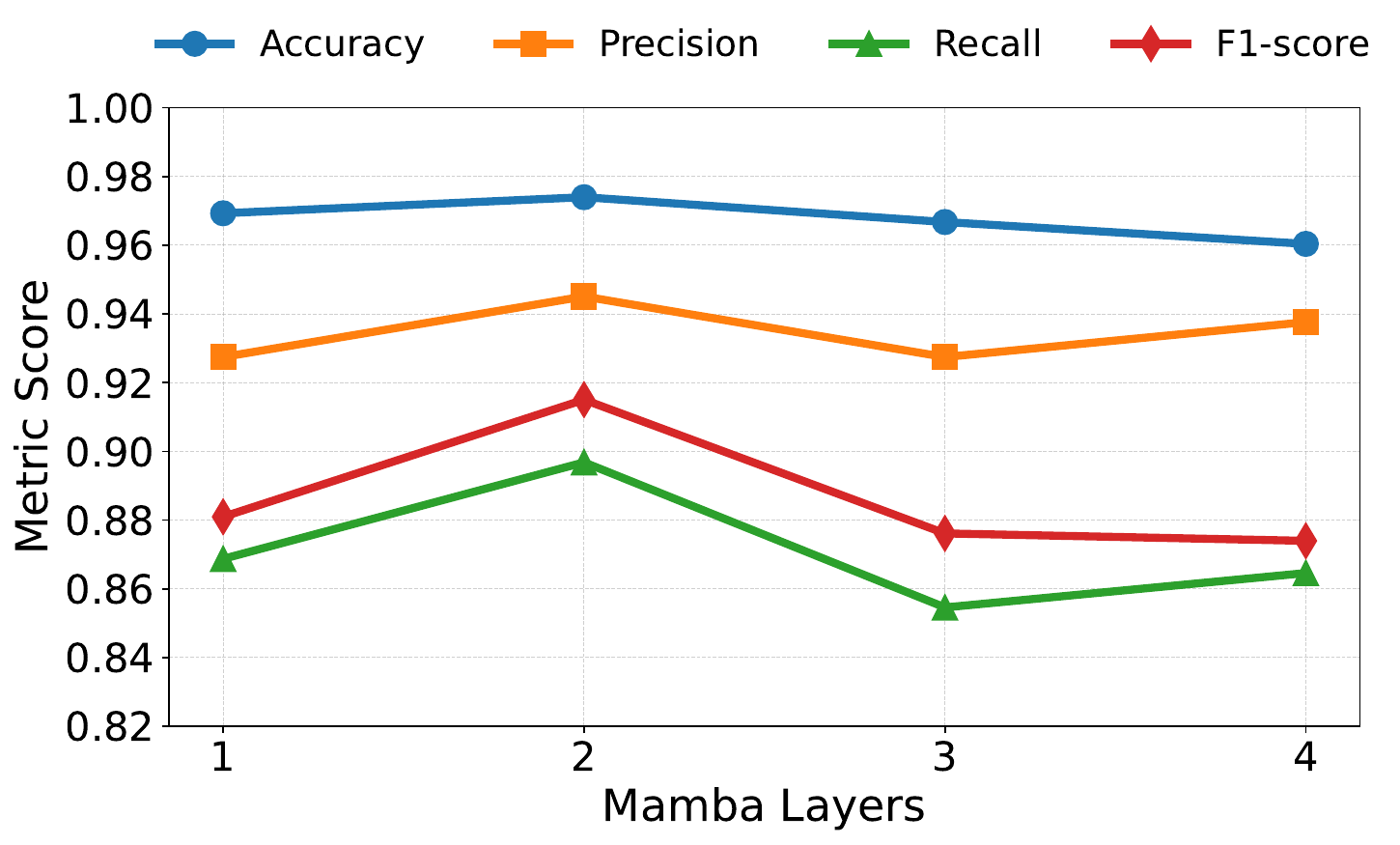}
    \par\vspace{2pt}
    (c) Effect of Mamba Encoder Layer Depth
  \end{minipage}
  \hspace{0.05\linewidth}
  \begin{minipage}[b]{0.43\linewidth}
    \centering
    \includegraphics[width=\linewidth]{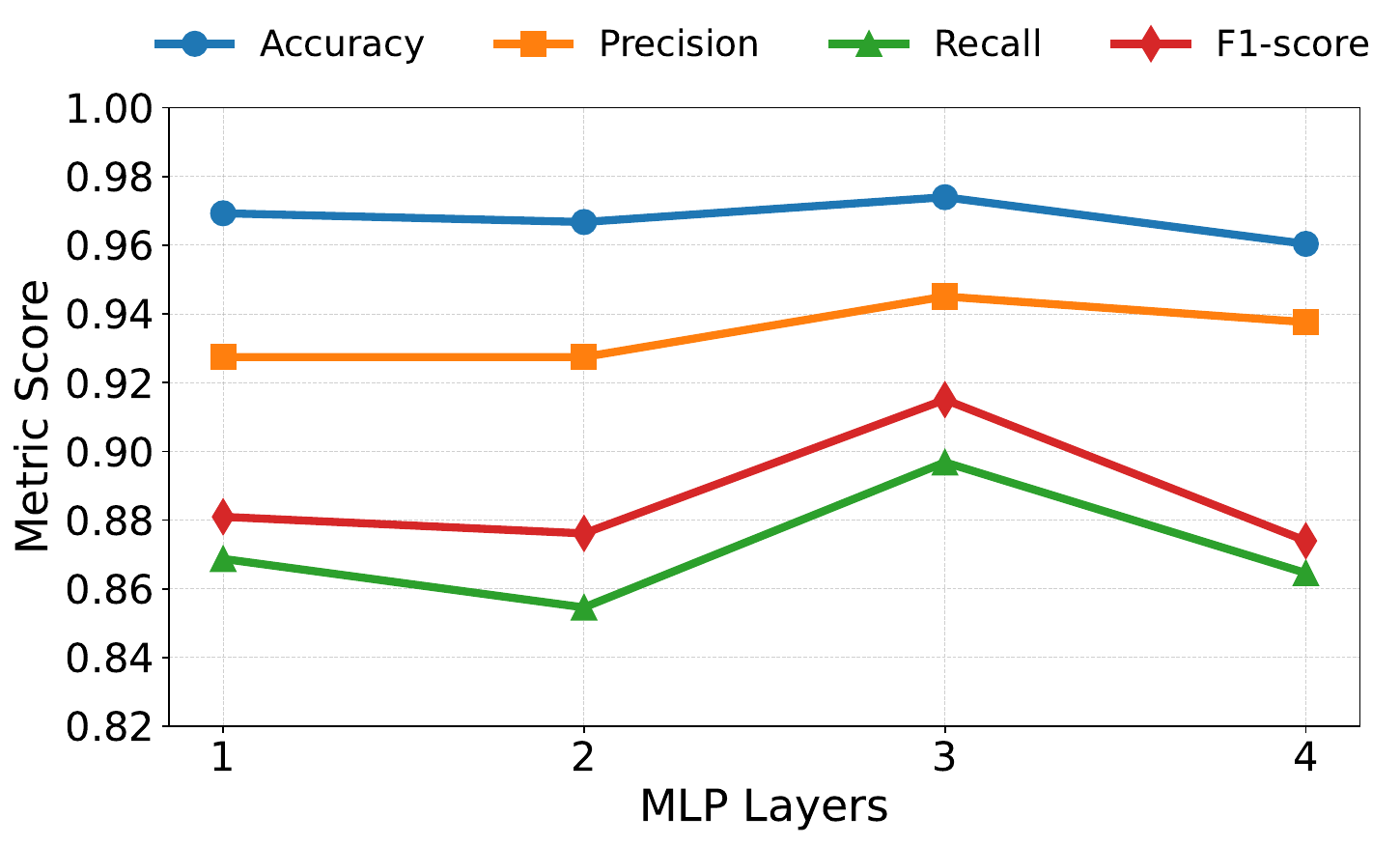}
    \par\vspace{2pt}
    (d) Effect of MLP Classifier Layer Depth
  \end{minipage}

  \caption{Impact of key hyperparameters on model performance: MVMD decomposition settings ($K$, $\alpha$), Mamba encoder depth, and MLP classifier depth.}
  \label{fig:model_architecture}
\end{figure*}

\subsection{Ablation Study}
As shown in Table~\ref{tab:ablation_minimal}, we conduct an ablation study on the CERT r4.2 dataset to evaluate the individual contributions of each core component in the Log2Sig framework. The full model integrates behavior encoding via the Mamba sequence encoder and multiscale decomposition through MVMD. This configuration achieves the best overall performance, with an F1-score of 0.956 across all evaluation metrics.

To assess component-wise impact, we test several reduced variants. Removing the Mamba encoder (w/o Mamba) while retaining MVMD leads to a moderate decline in performance (F1 = 0.929), indicating the importance of intra-day sequential modeling.
Excluding the frequency decomposition module (w/o MVMD) results in a larger decrease in recall and F1-score, highlighting the role of multiscale signal modeling. When both components are omitted (w/o MVMD and Mamba), the model exhibits the weakest performance (F1 = 0.856), confirming that both components are essential for robust anomaly detection.

\subsection{Parameter Sensitivity Analysis}
\label{sec:parameter}

\paragraph{MVMD Mode Number $K$}
As shown in Fig.~\ref{fig:model_architecture}(a), increasing the number of decomposition modes generally improves accuracy and precision, with performance peaking at $K=7$. This indicates that moderate multiscale resolution enhances the model’s ability to capture meaningful frequency components. However, when $K$ exceeds 10 (e.g., $K=13$ or $15$), performance degrades, particularly in recall and F1-score, likely due to the introduction of redundant or noisy modes. Based on this observation, $K=7$ is selected as the optimal setting.

\paragraph{MVMD Bandwidth $\alpha$}
Fig.~\ref{fig:model_architecture}(b) presents the sensitivity to the bandwidth parameter $\alpha$. The model remains relatively stable in the range of 300 to 1000, but larger values (e.g., $\alpha \geq 2000$) cause noticeable drops in recall. This degradation may stem from excessive smoothing, which reduces decomposition fidelity. A value of $\alpha=500$ is therefore adopted as it offers a robust trade-off between precision and generalization.

\paragraph{Mamba Encoder Layers}
According to Fig.~\ref{fig:model_architecture}(c), using 2 layers in the Mamba encoder achieves the most balanced performance. This depth is sufficient to model temporal dependencies while avoiding overfitting. Deeper configurations (3 or 4 layers) lead to performance declines, suggesting that additional layers may introduce overparameterization or vanishing gradients. Thus, a 2-layer encoder is used in our final setup.

\paragraph{MLP Classifier Layers}
As depicted in Fig.~\ref{fig:model_architecture}(d), increasing the depth of the MLP classifier enhances performance up to 3 layers, particularly in terms of recall and F1-score. However, further deepening to 4 layers results in performance drops, likely due to training instability or overfitting in the final classification stage. A 3-layer MLP is therefore adopted for the final architecture.

\begin{table}[!t]
\centering
\caption{Comparison of Multichannel Decomposition Methods (Per User Average).}
\label{tab:decomp_compare}
\begin{tabular}{lcccc}
\toprule
\textbf{Method} & \textbf{Acc} & \textbf{F1} & \textbf{Memory (MB)} & \textbf{Time (s)} \\
\midrule
MEMD & 0.965 & 0.911 & \textbf{1.17} & 6.37 \\
MVMD & \textbf{0.984} & \textbf{0.928} & 9.89 & \textbf{0.49} \\
\bottomrule
\end{tabular}
\end{table}

\begin{table}[!t]
\centering
\caption{Comparison of Sequence Encoder Methods (Per User Average).}
\label{tab:encoder_compare}
\begin{tabular}{lcccc}
\toprule
\textbf{Method} & \textbf{Acc} & \textbf{F1} & \textbf{GPU(MB)} & \textbf{Time (s)} \\
\midrule
LSTM        & 0.947             & 0.897             & 11.24             & 6.67 \\
Transformer & 0.963    & 0.912    & 8.89              & \textbf{6.11} \\
Mamba       & \textbf{0.981}    & \textbf{0.931}    & \textbf{6.33}            & \textbf{4.67} \\
\bottomrule
\end{tabular}
\end{table}

\subsection{Comparison of Decomposition and Encoder Strategies}

Table~\ref{tab:decomp_compare} and Table~\ref{tab:encoder_compare} summarize the performance and efficiency of different multichannel decomposition and sequence encoding strategies under a unified classification pipeline. All methods are evaluated with consistent preprocessing and training configurations to ensure fair comparison.

For decomposition, MVMD~\cite{ur2019multivariate} outperforms MEMD~\cite{rehman2010multivariate} with higher accuracy (0.984 vs. 0.965), F1-score (0.928 vs. 0.911), and dramatically lower computation time (0.49s vs. 6.37s), at the cost of slightly increased memory usage (9.89 MB vs. 1.17 MB). Note that both memory and time measurements refer solely to the decomposition stage, excluding downstream processing.

In terms of sequence encoding, Mamba achieves the best results across all metrics, with the highest accuracy (0.981) and F1-score (0.931), while also being the most efficient—consuming the least GPU memory (6.33 MB) and achieving the fastest inference speed (4.67s per user). These measurements are isolated to the encoder stage during per-user sequence modeling, excluding input preprocessing or classification layers.

In summary, the experimental results consistently demonstrate that MVMD and Mamba are the most effective and efficient components within their respective modules. MVMD significantly accelerates the multichannel decomposition process while improving classification performance, making it well-suited for real-time applications. Similarly, the Mamba encoder not only surpasses LSTM and Transformer in predictive accuracy but also offers superior computational efficiency with reduced GPU memory usage and inference time.

\section{Conclusion}
\label{sec:conclusion}
In this work, we proposed Log2Sig, a novel frequency-aware framework for insider threat detection that combines multivariate signal decomposition with deep sequence modeling. Unlike traditional methods that treat user logs as flat event sequences, Log2Sig transforms multichannel behavioral data into temporal signals and applies MVMD to extract frequency-localized intrinsic mode functions. These decomposed components, when fused with daily behavior statistics, reveal subtle and multiscale variations often overlooked by conventional models.
We further design a hybrid encoding architecture, where the daily behavior sequence is processed by a lightweight Mamba-based temporal encoder to capture long-range dependencies, while frequency-derived components are embedded and directly fused at the feature level. This enables efficient integration of semantic and spectral behavior cues for accurate anomaly detection.
Experiments on the CERT r4.2 and r5.2 datasets show that Log2Sig achieves consistently strong performance across different versions of the dataset.

\bibliographystyle{ieeetr}

\end{document}